\documentclass[aps,prd,twocolumn,amsmath,amssymb,nofootinbib]{revtex4-2}

\usepackage{bm}
\usepackage{graphicx}
\usepackage{hyperref}
\usepackage{xcolor}

\begin{document}

\title{Eckart heat-flux applicability in $F(\Phi,X)R$ theories and the existence of temperature gradients}

\author{David S. Pereira}%
	\email{djpereira@fc.ul.pt}
	\affiliation{%
		Departamento de F\'{i}sica, Faculdade de Ci\^{e}ncias da Universidade de Lisboa, Campo Grande, Edif\'{\i}cio C8, P-1749-016 Lisbon, Portugal}
	\affiliation{Instituto de Astrof\'{\i}sica e Ci\^{e}ncias do Espa\c{c}o, Faculdade de
		Ci\^encias da Universidade de Lisboa, Campo Grande, Edif\'{\i}cio C8,
		P-1749-016 Lisbon, Portugal
	}%

\author{José Pedro Mimoso}%
	\email{jpmimoso@fc.ul.pt}
	\affiliation{%
		Departamento de F\'{i}sica, Faculdade de Ci\^{e}ncias da Universidade de Lisboa, Campo Grande, Edif\'{\i}cio C8, P-1749-016 Lisbon, Portugal}
	\affiliation{Instituto de Astrof\'{\i}sica e Ci\^{e}ncias do Espa\c{c}o, Faculdade de
		Ci\^encias da Universidade de Lisboa, Campo Grande, Edif\'{\i}cio C8,
		P-1749-016 Lisbon, Portugal
	}%

\date{\today}

\begin{abstract}
We show that in single--scalar theories of the form $\mathcal{L}=F(\Phi,X)R+G(\Phi,X)$, a generic nonminimal coupling $F(\Phi,X)$ induces, in the scalar--comoving frame, an additional transverse contribution to the effective heat flux, proportional to $(F_X/8\pi F)V_{\perp a}$, where $V_a \equiv h_a{}^c\nabla_c\nabla_d X\,u^d$ and $V_{\perp a}$ denotes the component orthogonal to the 4--acceleration $a_a$. This term cannot in general be written as a spatial temperature gradient, and therefore obstructs a standard Eckart interpretation of the scalar sector for arbitrary timelike scalar configurations. As a result, requiring an Eckart heat flux $q_a = -K\bigl(D_a T_g + T_g\, a_a\bigr)$ for all such configurations is possible if and only if $F_X(\Phi,X)\equiv 0$, i.e.\ $F(\Phi,X)=F(\Phi)$, resulting in a theory that is a subclass of Horndeski. Thus, only Jordan--like theories of the type $F(\Phi)R+G(\Phi,X)$ admit a global Eckart fluid picture of the scalar sector, while models with $F_X\neq 0$ can recover an Eckart--like form only on highly symmetric backgrounds where the transverse contribution vanishes or collapses to a single gradient direction. We also make a brief comment on the existence of temperature gradients $D_aT_g$. 
\end{abstract}

\maketitle

\section{Introduction}

The theory of General Relativity (GR) \cite{Mimoso:2021sic,CANTATA:2021asi} is strongly intertwined with thermodynamics. The Hawking-Bekenstein entropy and temperature of black holes provide a remarkable and  widely appreciated milestone of this connection \cite{Bardeen:1973gs,Bekenstein:1973ur,Hawking:1975vc,Gibbons:1977mu,Visser:2003pp}. Yet the first studies of the thermodynamical features of GR date back to the work of  Tolman and Ehrenfest \cite{PhysRev.36.1791,Santiago:2018kds}  who addressed the thermal equilibrium in a static gravitational field. They have analysed the concept of temperature, associating it to a temperature gradient that prevents net heat flow. This offered  one of the first indications that, in curved spacetime, ``equilibrium'' is naturally phrased in terms of currents and their conservation (or non-conservation), rather than in terms of globally uniform intensive variables. 

Once dissipation is envisaged, the fluid viewpoint becomes more delicate. The original first-order theories of relativistic irreversible thermodynamics were associated with a choice of frame in  the works of Eckart \cite{Eckart:1940te}. A ``particle frame'' associated with the four-velocity congruence was tied to the particle current, while in the Landau--Lifshitz ``energy frame'' it was tied to the energy flow \cite{LandauLifshitz1987}. Although these formulations capture viscosity and heat conduction with simple constitutive laws, they are known to suffer from causality problems and, generically, from instabilities in the corresponding initial-value problem \cite{HiscockLindblom1985,Jou:1988pp,Pavon:1991pp,Maartens:1996vi}. The modern resolution of this issue arose from enlarging the set of dynamical variables, and moving to second-order, causal theories in the sense of M\"uller--Israel--Stewart (MIS), where dissipative fluxes relax on finite timescales, and the resulting equations are hyperbolic \cite{Muller1967,IsraelStewart1979}.

More generally, the identification of the various components in the energy-momentum tensor was carried in the seminal works by J. Ehlers \cite{Ehlers:2011aa} and  G. F. R. Ellis~\cite{Ellis:1971pg, Ellis:1973jva}. Further relevant aspects of the thermodynamics of closed and open gravitational systems have been intensively investigated in the suite~\cite{Coley:1986tt,Prigogine:1989aa,Gunzig:1987aa,Pavon:1988in,Pavon:1990ha,Lima:1990cj,Calvao:1991wg,Lima:1995kx,Lima:1995xz,Lima:1995kt,Carrillo:1996yu,Zimdahl:1996cg,Zimdahl:2000hq,Lima:2001lgd,Martin-Moruno:2013wfa,Martin-Moruno:2015bda,Martin-Moruno:2017exc,Mimoso:2016jwg,Mimoso:2018juu,Pereira:2024vdk}.

These works have disclosed the thermodynamics of the matter source terms of the Einstein field equations, and hence have mostly characterized the role of the right-hand side of the latter equations. 

However, since Jacobson's derivation of Einstein's equations from a local Clausius relation across Rindler horizons \cite{Jacobson:1995ab} the geometrical sector of the theory, i.e., the left-hand side of the equations, also became a focal point of interest. Subsequent progress made by Padmanabhan \cite{Padmanabhan:2009vy} has further strengthen the interpretation of  the field equations as a formulation of the first law, like identities, $T\, \mathrm{d}S=\mathrm{d}E+P\, \mathrm{d}V$. In this context, it was advocated that gravity should be understood as an emergent scenario in attempts to specify the gravitational entropy, which treat spacetime dynamics as the macroscopic limit of underlying statistical degrees of freedom
\cite{Verlinde:2010hp,Verlinde:2016toy,Clifton:2013dha,Acquaviva:2014owa}. 

The latter formulation is not an analogy, but, to some extent, depends on a judicious choice of variables: one decomposes the sources with respect to a timelike congruence, and the resulting kinematical and dynamical equations take the form of balance laws for energy, momentum, and entropy \cite{Ellis:1989jt,vanElst:1995eg,Ellis:1998ct,Ellis:2011hk}.

In a general manner, the starting point for a thermodynamic study is to consider a general dissipative decomposition of a symmetric tensor $T_{ab}$ relative to a unit timelike vector $u^a$ (with projector $h_{ab}=g_{ab}+u_a u_b$):
\begin{equation}\label{eq:general Tmunu}
T_{ab} = \rho\, u_a u_b + (P+\Pi)\, h_{ab} + q_a u_b + q_b u_a + \pi_{ab}\,,
\end{equation}
where $\rho=T_{ab}u^a u^b$ is the local energy density, $P$ the equilibrium pressure, $\Pi$ an effective bulk viscous pressure, $q_a$ the energy (heat) flux measured in the $u^a$-frame, and $\pi_{ab}$ the symmetric anisotropic stress (shear), with $q_a u^a=0$ and $\pi_{ab}u^b=0$. This decomposition isolates the covariant energy density, isotropic pressure, energy flow, and anisotropic stresses seen by the $u^a$-observers, thereby providing a convenient ``fluid'' bookkeeping quantity for interpreting any symmetric tensor---including effective sources arising from gravitational degrees of freedom~\cite{Espinosa-Portales:2021cac}. 

In modified gravity~\cite{Clifton:2011jh,Avelino:2016lpj,CANTATA:2021asi}, where there are extra degrees of freedom, this thermodynamic repackaging becomes especially important: one may shift all non-Einstein contributions to the right-hand side of the field equations and interpret them as an \emph{effective} stress--energy tensor for the additional gravitational sector~\cite{Harko:2014pqa,Capozziello:2014bqa,Capozziello:2013vna,Mimoso:2014ofa}.
For scalar--tensor theories, a particularly useful choice is the frame comoving with the scalar field---defined whenever the scalar gradient is timelike---in which the scalar sector can be written as an \emph{imperfect} relativistic fluid that has the form of Eq.~\eqref{eq:general Tmunu}. In Jordan-like models the associated effective heat flux often aligns with the four-acceleration and can be cast in an Eckart-type first-order form \cite{Faraoni:2018qdr,Faraoni:2021lfc,Faraoni:2025alq,Faraoni:2025fjq}. 
This imperfect-fluid construction has been developed systematically across broad classes of theories, including general scalar--tensor models, $f(R)$ gravity, and Horndeski and beyond--Horndeski sectors \cite{Faraoni:2018qdr,Giusti:2021sku,Miranda:2024dhw,Miranda:2022wkz,Faraoni:2025fjq}. 

A recent and particularly influential strand, due to Faraoni and collaborators \cite{Faraoni:2021jri,Giardino:2022sdv,Giusti:2022tgq,Faraoni:2023hwu}, has gone further by importing Eckart's first-order thermodynamics into this effective description.  By matching the scalar-fluid energy flow to Eckart's generalized Fourier law~\cite{Eckart:1940te}, one can identify a ``temperature of gravity'' and conductivity-like parameters directly from the covariant field reformulation, and use them to analyze the relaxation toward general relativity in homogeneous cosmologies~\cite{Damour:1992kf,Damour:1993id,Mimoso:1998dn,Mimoso:1999ai,Mimoso:2003iha}, as well as extensions to settings with past-directed timelike scalar gradients and anisotropic (Bianchi) universes. These developments make clear that an ``Eckart interpretation'' is not merely a change of language: it imposes a definite directional structure on the effective heat flux in the scalar-comoving frame. It is therefore important to understand \emph{when} such a standard Eckart interpretation is genuinely available, and what features of the theory obstruct it. The question of when an Eckart interpretation is available has been addressed systematically for Horndeski gravity in~\cite{Giusti:2021sku}.
There it is shown that the Eckart thermodynamic analogy does not apply to the most general Horndeski theory because certain operators spoil the constitutive relation for the effective scalar-fluid heat flux. This work further shows that the situation can be restored upon restricting the Horndeski action by removing the operator structures associated with intrinsic modifications of the helicity-2 sector (notably those tied to the gravitational-wave speed constraint), thereby recovering a standard Eckart form within a viable subclass.

Our aim here is to complement this general Horndeski picture by isolating, in a transparent and purely kinematical way, the obstruction mechanism associated with the single Effective Field theory (EFT) operator family $F(\Phi,X)R$. We address the question of whether the Eckart formalism is applicable to the class of metric single-scalar theories
\begin{equation}\label{eq:action}
S=\frac{1}{16\pi}\int d^4x\sqrt{-g}\,\Bigl[F(\Phi,X)R+G(\Phi,X)\Bigr]+S_m\,,
\end{equation}
where the scalar appears only through the pair $(\Phi,X)$ with
\begin{equation}
X\equiv-\frac12\nabla_a\Phi\nabla^a\Phi>0\,,
\end{equation}
for timelike scalar gradients. Allowing $F$ to depend on $X$ promotes the gravitational coupling
(or effective Planck mass) to a local quantity, $M_{\rm eff}^2(\Phi,X)\propto F(\Phi,X)$, and provides
a minimal covariant parametrization of kinetic mixing between the scalar and curvature. This
generalization is well motivated from the effective-field-theory viewpoint: in the unitary-gauge EFT
of inflation and dark energy the coefficient of $R$ is generically time dependent, and restoring
covariance via the St\"uckelberg construction naturally organizes operators into functions of
$(\Phi,X)$, leading to an effective $F(\Phi,X)R$ at the covariant level
\cite{Gubitosi:2012hu,Bloomfield:2012ff,Gleyzes:2013ooa,Frusciante:2019xia}. It is useful to regard
\eqref{eq:action} as an EFT truncation: in second-order or fully degenerate completions (such as
Horndeski and DHOST) a term of the form $F(\Phi,X)R$ is typically accompanied by additional operators
required to maintain the desired constraint structure~\cite{Kobayashi:2019hrl}. Moreover, invertible
disformal transformations generically generate $X$-dependent nonminimal couplings starting from
Jordan-like theories, reflecting the ubiquity of such kinetic mixing in general scalar--tensor
frameworks~\cite{Bettoni:2013diz,Zumalacarregui:2013pma,Bettoni:2015wta}. The family~\eqref{eq:action}
therefore interpolates between standard Jordan-frame nonminimal couplings, recovered when
$F(\Phi,X)=F(\Phi)$ (including metric $f(R)$ in its scalar representation), and broader kinetic
nonminimal couplings with $F_X\neq 0$ frequently used in EFT-inspired model building.

The main point of this paper is to show that this generalization has a sharp and purely kinematical consequence for the effective-fluid interpretation. Working in the scalar-comoving frame, we show that an $X$-dependent nonminimal coupling generically induces an \emph{additional} contribution to the scalar-sector heat-flux $q_a^{(\Phi)}$ that is not aligned with the four-acceleration $a_a$. Crucially, this new piece cannot in general be expressed as a spatial gradient, and therefore cannot be consistently absorbed into the Eckart heat-flux term $q_a=-K\bigl(D_aT_g+T_g a_a\bigr)$~\cite{Eckart:1940te} for arbitrary configurations. As a result, demanding that the scalar sector admit a standard Eckart interpretation in its own comoving frame for \emph{all} timelike scalar configurations yields a simple selection rule: it is possible if and only if $F_X(\Phi,X)\equiv 0$, i.e.\ $F(\Phi,X)=F(\Phi)$. This motivates a practical admissibility criterion for scalar--curvature couplings phrased directly in the \emph{directional structure} of the scalar-sector heat flux in the scalar-comoving frame. We also comment on how temperature-gradient terms should be understood once one distinguishes the components parallel and orthogonal to the
acceleration.

Our analysis is off shell and kinematical: we do not assume any particular background or
use the field equations beyond the definition of the effective stress--energy tensor. In this sense, we isolate the distinctive imprint of kinetic nonminimal couplings $q_{a}^{(\Phi)}$, and clarify why highly symmetric configurations can accidentally hide the obstruction (because the transverse direction may vanish or collapse to a
single gradient direction).

The paper is organized as follows. In Sec.~\ref{sec:II} we introduce the scalar-comoving frame and the $1+3$ kinematical identities that control the available spatial directions built from a single timelike scalar. In Sec.~\ref{sec:III} we briefly revisit the Eckart-form heat flux for Jordan-like $F(\Phi)R$ models and emphasize the role of the parallel/perpendicular decomposition of temperature gradients. In Sec.~\ref{sec:IV} we derive the general heat flux for~\eqref{eq:action}, identify the new transverse contribution proportional to $F_X$, and formulate the Eckart-based selection rule. In Sec.~\ref{sec:V} we provide an example where the new contribution cannot be written as a gradient of a scalar and illustrate how high symmetry (e.g.\ Friedmann-Lema\^itre-Robertson-Walker (FLRW) or spherical symmetry) can eliminate the transverse obstruction even when
$F_X\neq 0$. We conclude with a discussion and outlook.

\section{Scalar-comoving frame and $1+3$ kinematics}\label{sec:II}

We assume throughout that the scalar gradient is timelike and define, using the standard EFT convention for $X$,
\begin{equation}
  u_a \equiv \frac{\nabla_a\Phi}{\sqrt{2X}},\quad u_a u^a=-1.
\end{equation}

The corresponding spatial projector is
$h_{ab}\equiv g_{ab}+u_a u_b$, with $h_a{}^b u_b=0$.
Since $u_a\propto\nabla_a\Phi$ is hypersurface-orthogonal, its vorticity vanishes and the derivative of $u_a$ admits the standard decomposition
\begin{equation}\label{eq: u decomposed}
  \nabla_a u_b = - u_a a_b + \frac{1}{3}\theta h_{ab} + \sigma_{ab},
\end{equation}
where $a_a\equiv u^b\nabla_b u_a$ is the four-acceleration, $\theta\equiv \nabla_a u^a$ the expansion, and $\sigma_{ab}$ the shear tensor (symmetric, trace-free, and orthogonal to $u^a$).
We denote spatial derivatives by
$D_a \ () \equiv h_a{}^b\nabla_b \ ()$ .

Two kinematical identities play a central role in what follows.
First,
\begin{equation}
  D_a X = -2X\,a_a,
  \label{eq:Dax}
\end{equation}
which follows directly from $X=-\tfrac12\nabla_c\Phi\nabla^c\Phi$ and the definition of $u_a$.
Second,
\begin{equation}
  h_a{}^c\nabla_c\nabla_d\Phi\,u^d = \sqrt{2X}\,a_a.
  \label{eq:keyid}
\end{equation}

Equations~\eqref{eq:Dax} and~\eqref{eq:keyid} are equivalent and express the fact that the only independent spatial vector built from first and second derivatives of a single timelike scalar is the acceleration $a_a$.

\section{Eckart heat-flux for $F(\Phi)R$}\label{sec:III}

Prior to analyzing the $F(\Phi,X)$ case, we will revisit the effective-fluid framework introduced by Faraoni and collaborators in Refs.~\cite{Faraoni:2018qdr,Faraoni:2021lfc}. As mentioned in the introduction, the starting point is simple: After obtaining the field equations for the scalar-tensor theory, one rearranges the Einstein equation so that every term not belonging to the Einstein tensor is shifted to the right side and interpreted as an effective stress--energy tensor for the extra gravitational degrees of freedom. After choosing the comoving frame $u_a\propto\nabla_a\Phi$ and carrying out a $1+3$ splitting this effective tensor inherently assumes the structure of a \emph{imperfect} relativistic fluid. Specifically the heat flux is an exact covariant quantity derived from the kinematics and the nonminimal coupling. This perspective has shown to be a functional cohesive terminology throughout scalar-tensor gravity~\cite{Faraoni:2018qdr,Faraoni:2021lfc,Faraoni:2025alq,Faraoni:2025fjq} and associated frameworks and it specifies when and the manner, in which a thermodynamic interpretation à la Eckart becomes available in the comoving frame. Most importantly, it provides a thermodynamic (thermal) framework/analogy to interpret departures from GR in scalar–tensor gravity (and related cases where the formalism applies) as finite-‘temperature’ deviations from the GR (‘zero-temperature’) state, as done in~\cite{Faraoni:2025alq}.

To make these statements concrete, we start from the Jordan-frame scalar--tensor action~\cite{Faraoni:2004pi}
\begin{equation}
  S = \frac{1}{16\pi}\int d^4x\sqrt{-g}\,
  \bigl[F(\Phi)R + Z(\Phi)X - 2V(\Phi)\bigr] + S_m.
  \label{eq:JFST}
\end{equation}

The modified Einstein equation can be written as
\begin{equation}
  F(\Phi)G_{ab} = 8\pi T^{(m)}_{ab} + T^{(\Phi)}_{ab},
\end{equation}
with the scalar-sector stress--energy tensor
\begin{align}
  T^{(\Phi)}_{ab} &=
   \nabla_a\nabla_b F - g_{ab}\Box F
\nonumber\\
  &\quad - Z(\Phi)\Bigl(\nabla_a\Phi\nabla_b\Phi - \tfrac{1}{2}g_{ab}\nabla_c\Phi\nabla^c\Phi\Bigr)
   - g_{ab}V(\Phi).
\end{align}
We interpret $T^{(\Phi)}_{ab}$ as an effective fluid and define
$T^{\mathrm{eff}}_{ab}\equiv T^{(\Phi)}_{ab}/(8\pi F)$, so that
\begin{equation}
  G_{ab} = 8\pi \frac{T^{(m)}_{ab}}{F(\Phi)} + 8\pi T^{\mathrm{eff}}_{ab}.
\end{equation}

In the scalar-comoving frame, the effective heat flux is
\begin{equation}
  q^{(\Phi)}_a \equiv -h_a{}^c T^{\mathrm{eff}}_{cd}u^d.
\end{equation}

Terms proportional to $g_{ab}$ or to $\nabla_a\Phi\nabla_b\Phi\propto u_a u_b$ do not contribute to $q^{(\Phi)}_a$, because $h_a{}^c u_c=0$.
Thus
\begin{equation}
  q^{(\Phi)}_a = -\frac{1}{8\pi F}\,h_a{}^c\nabla_c\nabla_d F\,u^d.
\end{equation}

For $F=F(\Phi)$ one has
$\nabla_a\nabla_b F = F'(\Phi)\nabla_a\nabla_b\Phi + F''(\Phi)\nabla_a\Phi\nabla_b\Phi$, and again the second term drops out of the heat flux.
Using~\eqref{eq:keyid} we obtain
\begin{equation}
  q^{(\Phi)}_a =
  -\frac{F'(\Phi)}{8\pi F(\Phi)}\sqrt{2X}\,a_a.
  \label{eq:q_Fphi}
\end{equation}

Considering now Eckart's first-order constitutive heat flux relation~\cite{Eckart:1940te} 
\begin{equation}\label{eq:eckart1st}
  q_a = -K\bigl(D_a T_g + T_g a_a\bigr),
\end{equation}
with conductivity $K$ and temperature $T_g$, one can compare the $q_a^{(\Phi)}$ with Eckart's first law and extract thermal relations for $T_g$. In the literature, usually $D_a T_g$ is set to zero, or inferred as zero from Eq.~\eqref{eq:q_Fphi}, allowing to identify the invariant product
\begin{equation}
  K T_g = \frac{F'(\Phi)}{8\pi F(\Phi)}\sqrt{2X}.
\end{equation}

Only the product $K T_g$ is fixed; rescalings $K\to\beta K$, $T_g\to T_g/\beta$ leave the stress--energy tensor unchanged.
Equation~\eqref{eq:q_Fphi} reproduces, up to conventions for $X$, the expressions obtained in scalar--tensor and metric $f(R)$ thermodynamic analyses~\cite{Faraoni:2018qdr,Faraoni:2021lfc}.

However, a more detailed analysis can be done. Reading $D_a T = 0$ from Eq.~\eqref{eq:q_Fphi} must be done with some care. Due to $D_a T$ and $a_a $ being spatial vectors, i.e, $u^aD_a T = 0$ and $u^a a_a=0$, it is possible to decompose $D_aT$ into a part parallel and
a orthogonal to $a_a$ in the comoving 3--space. We write
\begin{equation}
  D_a T = (D_a T)_\parallel + (D_a T)_\perp,
\end{equation}
where
\begin{equation}
  (D_a T_g)_\parallel \equiv \lambda\,a_a,
  \qquad
  \lambda \equiv \frac{D_b T_g\,a^b}{a^2},
\end{equation}
and so
\begin{equation}
  (D_a T_g)_\perp \equiv D_a T_g - (D_a T_g)_\parallel,
  \qquad
  (D_a T_g)_\perp a^a = 0.
\end{equation}

This decomposition allows one to understand that in the general, because to have $q_a^{(\Phi)} \propto a_a$ then $(D_a T)_{\perp}=0$ but $(D_a T)_\parallel \neq0$. The condition $D_a T = 0$ only truly applies to specific symmetric configurations such as FLRW. 

Moreover, by introducing the projector onto the 2D subspace orthogonal to
$a_a$ (with $a_a\neq0$) inside the spatial 3--space,
\begin{equation}\label{eq:Projetor}
  P_a{}^b \equiv h_a{}^b - \frac{a_a a^b}{a^2},
\end{equation}
we have
\begin{equation}
  (D_a T)_\perp = P_a{}^b D_b T,
  \qquad
  (D_a T)_\parallel = D_a T - (D_a T)_\perp\,,
\end{equation}
a decomposition that will be used below to study the $F(\Phi,X)$ case. 

\section{Eckart Heat-flux for general $F(\Phi,X)R+G(\Phi,X)$ theories}\label{sec:IV}

We now return to the general action~\eqref{eq:action}.
Varying the $F(\Phi,X)R$ term yields
\begin{equation}\label{eq:STGeneral}
  T^{(\Phi)}_{ab} =\left(\nabla_a\nabla_b - g_{ab}\Box\right)F+\Delta\nabla_a\Phi\nabla_b\Phi+ \frac12g_{ab}G\,,
\end{equation}
where $\Delta=\frac12\bigl(R F_X + G_X\bigr)$. Defining now an effective energy--momentum tensor
\[
  T^{\mathrm{eff}}_{ab} \equiv \frac{T^{(\Phi)}_{ab}}{8\pi F},
\]
the corresponding energy flux is given by
\begin{equation}
  q^{(\Phi)}_a \equiv -h_a{}^c T^{\mathrm{eff}}_{cd} u^d
  = -\frac{1}{8\pi F}\,h_a{}^c \nabla_c \nabla_d F\,u^d\,,
  \label{eq:q_FphiX_def}
\end{equation}
where the only term from Eq.~\eqref{eq:STGeneral} that survived was $\nabla_a\nabla_b F$ because all the other terms are either proportional to $g_{ab}$ or to $u_a u_b$ and cannot contribute to $q^{(\Phi)}_a$ as
\begin{equation}\label{eq:fund results h u}
  h_a{}^c g_{cd}u^d = h_a{}^c u_c = 0,\qquad
  h_a{}^c u_c u_d u^d = 0.
\end{equation}

Expanding then $\nabla_c\nabla_d F(\Phi,X)$ via the chain rule yields
\begin{align}
  q^{(\Phi)}_a
  &= -\frac{1}{8\pi F}\,h_a{}^c\Big[
    F_\Phi \nabla_c\nabla_d\Phi
   + F_{\Phi\Phi}\,\nabla_c\Phi\nabla_d\Phi
\nonumber\\
  &\qquad\qquad\quad
   + F_{\Phi X}\,\nabla_c X\nabla_d\Phi
   + F_X\nabla_c\nabla_d X
\nonumber\\
  &\qquad\qquad\quad
   + F_{X\Phi}\,\nabla_c\Phi\nabla_d X
   + F_{XX}\,\nabla_c X\nabla_d X
  \Big]u^d\,.
\end{align}

Using $h_a{}^c \nabla_c \Phi = 0$, $u^d \nabla_d \Phi = -\sqrt{2X}$, together with the kinematical identities~\eqref{eq:Dax} and~\eqref{eq:keyid}, we obtain
\begin{align}
  q^{(\Phi)}_a
  &= -\frac{1}{8\pi F}\Bigl[
    F_\Phi\sqrt{2X}
    + 2\sqrt{2X}\,X F_{\Phi X}
    - 2X\dot X F_{XX}
  \Bigr]a_a
\nonumber\\
  &\quad
  -\frac{1}{8\pi F}F_X V_a,
  \label{eq:q_FphiX_full_ped}
\end{align}
where $V_a$ is defined as
\begin{equation}
     V_a \equiv h_a{}^c\nabla_c\nabla_d X\,u^d.
\end{equation}
that arises from the term $\nabla_c \nabla_d X$ and is the only contribution in $\nabla_c\nabla_d F$ capable of producing a spatial vector that is not necessarily proportional to the acceleration $a_a$.

By making use of the $1+3$ decomposition of $\nabla_a u_b$~\eqref{eq: u decomposed} and the scalar kinematics, one can show that $V_a$ takes the general form
\begin{equation}
  V_a = D_a\dot X - \frac{1}{3}\theta D_a X - \sigma_a{}^b D_b X,
  \label{eq:Va_final}
\end{equation}
where $\dot X \equiv u^b \nabla_b X$ is the derivative of $X$ along $u^a$.

In this form, the relation of $V_a$ to the acceleration $a_a$ is not immediately apparent. However, using~\cite{Tsagas:2007yx}
\begin{equation}
    D_a \dot{X} - h_a{}^b \dot{\left(D_b X\right)} = - \dot{X} a_a + \frac{1}{3} \theta D_a X + \sigma_a{}^b D_b X \,,
\end{equation}
the expression for $V_a$ can be simplified to
\begin{equation}
    V_a = -3 \dot{X} a_a - 2X\,h_a{}^b \dot{a}_b\,,
\end{equation}
which makes explicit that $V_a$ contains a component proportional to $a_a$ as well as, in general, an additional component, a jerk, that may not be only proportional to $a_a$. So, in a general way one can decompose $V_a$ as
\begin{equation}
  V_a = V_{\parallel} a_a + V_{\perp a},
\end{equation}
where $V_{\perp a} = P^{\ b}_aV_b = -2X j_{\perp a}$ with  $j_{\perp a} = P^{\ b}_a h_{b}^{\ c} \dot{a}_c$. 
Collecting the contributions along $a_a$ into a scalar coefficient $f(\Phi,X,\dot X,\ldots)$, the scalar heat flux can therefore be written as
\begin{equation}
  q^{(\Phi)}_a
  = -f(\Phi,X,\dot X,\ldots)\,a_a
    - \frac{F_X}{8\pi F}\,V_{\perp a}.
  \label{eq:q_decomp}
\end{equation}

The first term corresponds to an ``Eckart-like'' contribution, in which the heat flux is aligned with the fluid acceleration and can be interpreted as a thermal--acceleration piece. By contrast, the second term selects an additional spatial direction in the scalar--comoving frame, which in general cannot be recast as a simple multiple of $a_a$.

This transverse contribution is absent in standard $F(\Phi)R$ scalar--tensor theories (for which $F_X=0$)~\cite{Faraoni:2018qdr}. It is also absent in the GW-compatible $c_T=1$ Horndeski sector (e.g.\ $G_{4X}=0$ and $G_5=0$), whose effective scalar-fluid heat flux aligns with $a_a$ in the comoving frame~\cite{Giusti:2021sku,Miranda:2022wkz,Miranda:2024dhw}. We stress, however, that outside such restricted subclasses the Eckart structure can fail already within Horndeski theory, as shown explicitly in~\cite{Giusti:2021sku}. In this sense, within the EFT-truncated family~\eqref{eq:action} the term proportional to $F_X V_{\perp a}$ provides a simple and explicit mechanism for generating a non-Eckart directional structure in $q^{(\Phi)}_a$.

\subsection{Selection rule from Eckart structure}

We start by decomposing the Eckart relation~\eqref{eq:eckart1st} into parts parallel and orthogonal to $a_a$,
\begin{equation}\label{eq:Eckart Decomposed}
    q_a = - K \left\{ (D_a T_g)_{\perp} + (T_g + \lambda)\,a_a \right\}\,.
\end{equation}

Comparing the previous expression with Eq.~\eqref{eq:q_decomp}, the part of
Eq.~\eqref{eq:q_decomp} proportional to $a_a$ can always be written in Eckart form, $-K (T_g + \lambda) a_a$, for a suitable choice of the scalar combination $T_g + \lambda$. The remaining term, proportional to $V_{\perp a}$, identifies an additional spatial direction in the comoving three--space. Although it is possible to consider that the Eckart conduction term, $-K (D_a T_g)_{\perp}$, can be equal to  $\frac{F_X}{8\pi F}\,V_{\perp a}$ this is only true if, for the configuration under consideration, $V_{\perp a}$ is the spatial gradient of a scalar potential in the comoving three--space; that is, if there exists a scalar field $\Psi$ such that $V_{\perp a} = D_a \Psi$, or equivalently, if $V_{\perp a}$ is irrotational in that three--space. However, this condition is not in general satisfied for inhomogeneous and shearing scalar flows. More precisely, computing $\epsilon_{abc} D^{b} V^c$ where $\varepsilon_{abcd} = \sqrt{h}\eta_{abc}$ with $\eta_{abc}$ being the Levi–Civita symbol, yields
\begin{equation}\label{eq:curl V perp}
    \varepsilon_{abc} D^b V_{\perp}^c = 4X\varepsilon_{abc} a^{b}j_{\perp}^c - 2X \varepsilon_{abc} D^b j_{\perp}^c\,,
\end{equation}
that showcases, in general, that $V_\perp$ is not irrotational and therefore, the application of the Eckart formalism for general configurations is not compatible with the existence of the term $\frac{F_X}{8\pi F}\,V_{\perp a}$. This result can be viewed as a way to constraint the theory. Eckart theory provides the minimal relativistic extension of Fourier's law: it is the standard first--order framework in which a relativistic fluid is endowed with a scalar temperature, a spatial temperature gradient \(D_aT\), and a heat flux \(q_a\) obeying Eq.~\eqref{eq:Eckart Decomposed}, thus tying together temperature gradients and acceleration in a way that is closely related to the Tolman--Ehrenfest relation and other aspects of gravitational thermodynamics (for more details on the connection of the Eckart constitutive law with the Tolman--Ehrenfest viewpoint, including an Einstein-frame derivation see~\cite{Karolinski:2024ukr}).

In our setting the natural four--velocity is the scalar--comoving one, \(u_a \propto \nabla_a\Phi\), and the corresponding heat flux \(q^{(\Phi)}_a\) captures the nontrivial energy transport induced by nonminimal couplings. Demanding that this effective scalar fluid admit a standard Eckart interpretation for all timelike configurations is therefore a very conservative requirement: it is the weakest and most familiar relativistic notion of ``having a temperature and a Fourier--like heat conduction'' in this frame. Requiring that the theory admits~\eqref{eq:Eckart Decomposed} for all configurations translates into imposing $\frac{F_X}{8\pi F}\,V_{\perp a} = 0$ that is only true \emph{for all configurations} if
\begin{equation}
  F_X(\Phi,X)\equiv 0 \quad\Rightarrow\quad F(\Phi,X)=F(\Phi),
\end{equation}
as it is possible for $V_{\perp a}\neq0$ for some configurations. This condition ensures that the transverse term in~\eqref{eq:q_decomp} vanishes, and considering $\left(D_aT\right)_{\perp }=0$ we recover the purely accelerative heat flux~\eqref{eq:q_Fphi} for all configurations. Remarkably, this condition restricts the theory to a subclass of Horndeski models with $G_3 = 0$ and satisfying the gravitational–wave constraints of~\cite{Baker:2017hug,Creminelli:2017sry}, namely $G_{4X}(\Phi,X) = 0$ and $G_5(\Phi,X) = 0$. Hence, this result shows that even this minimal demand already acts as a powerful model--building filter: it selects the subclass \(F(\Phi,X)=F(\Phi)\) out of the more general family \(F(\Phi,X)R+G(\Phi,X)\). In other words, only Jordan--like nonminimal couplings allow the scalar sector to behave as a single--temperature relativistic fluid with an Eckart heat flux in its own comoving frame; for \(F_X\neq 0\) the scalar sector exhibits a more general transport structure that cannot be captured by Eckart's law for general configurations.

With this result we can formulate the \emph{Proposition:} For metric single-scalar theories of the form~\eqref{eq:action} with timelike scalar gradient and no higher derivatives of $\Phi$ in the Lagrangian, the scalar sector admits a pure Eckart-like heat flux in the scalar-comoving frame, for arbitrary such configurations, if and only if $F(\Phi,X)$ is independent of $X$, i.e.\ $F_X(\Phi,X)\equiv 0$ and $F(\Phi,X)=F(\Phi)$. In that case the theory is of the $F(\Phi)R+G(\Phi,X)$ type, and the heat flux takes the form $q^{(\Phi)}_a = -f(\Phi,X,\dot X)a_a$ with an invariant product $K(\lambda + T_g)=f(\Phi,X,\dot X)$ for an appropriate choice of $K$ and $T_g$.

It is important to emphasize that imposing the condition $F_X(\Phi,X) = 0$ is to guarantee that the theory admits an Eckart heat-flux description for all configurations and if one desires to explore a specific sets of configurations this condition may be not necessary. In particular, on highly symmetric backgrounds---for example, homogeneous FLRW spacetimes or spherically symmetric configurations with $\Phi = \Phi(t,r)$ and timelike gradient (both of which will be analyzed in the following section)---the vector $V_{\perp a}$ may either vanish identically or be confined to a single radial direction that is itself a gradient. In such special cases, $V_{\perp a}$ can be absorbed into the function $f(\Phi,X,\dot{X})$, and the scalar sector admits an Eckart--like description even when $F_X \neq 0$, albeit only within this restricted class of configurations. Additionally, the configuration cases where $V_{\perp a} = D_a \Psi$ present a possibly new rich avenue to study the thermodynamical characteristics of gravity, specially scalar-tensor gravity. 

Therefore, within the family $\mathcal{L}=F(\Phi,X)R+G(\Phi,X)$, the effective heat flux in the scalar-comoving frame generically contains an additional component orthogonal to the 4-acceleration whenever $F_X\neq 0$. This result should be read in the light of the general Horndeski analysis of~\cite{Giusti:2021sku}: within Horndeski theory the Eckart constitutive structure is not available in full generality, but it can be recovered in restricted (phenomenologically viable) subclasses. Here we show that, within the EFT-truncated family $F(\Phi,X)R+G(\Phi,X)$, demanding a global Eckart interpretation for all timelike scalar configurations is equivalent to eliminating the unique source of a generically non-integrable transverse heat-flux direction, namely $F_X V_{\perp a}$. In particular, the condition $F_X=0$ removes the transverse contribution identically and reduces the scalar-comoving heat flux to the purely accelerative (Jordan-like) form.

A natural question concerns the frame-(in)dependence of the thermodynamic analogy. In this work we adopt the Jordan-frame viewpoint, in which the nonminimal coupling is explicit and the modified field equations can be recast as Einstein equations sourced by an \emph{effective} scalar stress--energy tensor that admits an imperfect-fluid decomposition in the scalar-comoving frame. For the standard scalar--tensor case with $F(\Phi)R$ coupling, a conformal transformation yields an Einstein-frame representation with an Einstein--Hilbert gravitational sector and a (redefined) minimally coupled scalar; the status of the fluid analogy and of the associated ``thermodynamic'' identifications in that conformal frame has been analyzed in detail in Ref.~\cite{Faraoni:2022gry}. By contrast, for the broader family considered here, $\mathcal{L}=F(\Phi,X)R+G(\Phi,X)$, an $X$-dependent nonminimal coupling does not in general admit a simple \emph{conformal} Einstein-frame reformulation: invertible field redefinitions that remove kinetic mixing typically involve more general (disformal/derivative-dependent) transformations and can be model dependent \cite{Bettoni:2013diz,Zumalacarregui:2013pma,Bettoni:2015wta}. For this reason, the criterion derived in this work should be read as a Jordan-frame statement about the existence of a \emph{standard Eckart} heat-flux interpretation of the effective scalar sector in the frame where the coupling $F(\Phi,X)R$ is manifest.

\section{Notable configurations}\label{sec:V}

\subsection{Stationary axisymmetric configuration with $\mathrm{curl}\,V_{\perp}\neq 0$}
\label{sec:notable:rotating}

Here we show a simple stationary axisymmetric configuration in which the scalar-comoving flow is accelerated
($a_a\neq 0$) and, crucially, the transverse vector $V_{\perp a}$ has nonvanishing 3D curl in the comoving
three-space: $\varepsilon_{abc}D^b V_{\perp}^{\,c}\neq 0$ that implies that
$V_{\perp a}\ \text{is not (locally) a spatial gradient in general.}$
This provides a concrete example where the direction selected by $V_{\perp a}$ cannot be identified with a
(comoving, spatial) temperature gradient.

We start by considering the stationary axisymmetric circular line element
\begin{equation}
\label{eq:notable:rot_metric}
ds^2
=
-N(r)^2\,dt^2
+B(r)\,dr^2
+r^2\,d\theta^2
+\psi^2\,\bigl(d\varphi-\omega(r)\,dt\bigr)^2,
\end{equation}
with $\psi=r\sin(\theta)$, $N(r)>0$ and $B(r)>0$.
(If one wishes to work strictly to first order in rotation, one may drop the $\omega^2$ term inside $g_{tt}$;
the final results below are linear in $\omega'(r)$.)

For the scalar-field we choose the timelike scalar profile
\begin{equation}
\label{eq:notable:rot_scalar}
\Phi = q\,t,\qquad q< 0\,,
\end{equation}
this gives
\begin{equation}
\label{eq:notable:rot_X_u}
X = \frac{q^2}{2N(r)^2}>0,
\end{equation}
and
\begin{equation}\label{eq:ua V}
u_a = -N(r)\,\nabla_a(t),
\end{equation}
and thus the scalar-comoving hypersurfaces are $t=\mathrm{const}$ and the induced 3-metric is
\begin{equation}
\label{eq:notable:rot_3metric}
h_{ij}=\mathrm{diag}\bigl(B(r),\,r^2,\,r^2\sin^2\theta\bigr),
\end{equation}
with the 3D Levi--Civita tensor normalized by
\begin{equation}
\label{eq:notable:rot_eps3}
\varepsilon_{r\theta\varphi}=+\sqrt{\det h}=+\,r^2\sin\theta\,\sqrt{B(r)}.
\end{equation}

Since $X$ depends only on $r$, the only nonzero spatial derivative is radial
\begin{equation}
D_r X = \partial_r X = \partial_r\!\left(\frac{q^2}{2N^2}\right)
= -\frac{q^2}{N^3}N'\,,
\end{equation}
allowing to conclude by using Eq.~\eqref{eq:Dax} that
\begin{equation}
a_r = -\frac{D_r X}{2X}
= \frac{N'}{N},
\qquad
a_\theta=a_\varphi=0.
\label{eq:a_cov}
\end{equation}

To compute $\varepsilon_{abc}D^b V_{\perp}^c$ it is necessary to first compute $\dot{a}_a$ as $j_{\perp a} \propto \dot{a}_a$. For the scalar–comoving congruence
$u^a=(u^t,0,0,u^\varphi)$ (i.e. $u^r=u^\theta=0$), the convective derivative of $a_r$ reads
\begin{align}
\dot a_r
&= u^b\partial_b a_r - u^b\Gamma^{c}{}_{br}a_c \\ \nonumber
&= u^t\partial_t a_r + u^\varphi\partial_\varphi a_r
- (u^t\Gamma^{r}{}_{tr}a_r+u^\varphi\Gamma^{r}{}_{\varphi r}a_r)\,,
\end{align}
where partial-derivative terms vanish because $\partial_t a_r=\partial_\varphi a_r=0$, and the connection sources also vanish
\begin{equation}
\Gamma^{r}{}_{tr}=\Gamma^{r}{}_{\varphi r}=0,
\end{equation}
since the metric is circular (no $g_{tr}$, $g_{r\varphi}$) and $g_{rr}$ is independent of $t,\varphi$.
Therefore $\dot a_r=0$. Similarly, since $a_\theta=0$ one has identically,
\begin{equation}
\dot a_\theta \equiv u^b\nabla_b a_\theta
= -u^b\Gamma^{c}{}_{b\theta}a_c
= -(u^t\Gamma^{r}{}_{t\theta}+u^\varphi\Gamma^{r}{}_{\varphi\theta})\,a_r,
\end{equation}
and the relevant mixed connections also vanish,
\begin{equation}
\Gamma^{r}{}_{t\theta}=\Gamma^{r}{}_{\varphi\theta}=0,
\end{equation}
because $g_{t\theta}=g_{\varphi\theta}=0$ and the metric has no $(t,\varphi)$--$(r,\theta)$ cross terms.
Hence $\dot a_\theta=0$.

For $a_\varphi$ one has
\begin{equation}
\label{eq:notable:rot_dotaphi_start}
\dot a_\varphi
=
u^b\nabla_b a_\varphi
=
-\,u^b\Gamma^{r}{}_{b\varphi}\,a_r.
\end{equation}
that in contrast with the previous cases the connections are not yielding
\begin{equation}
\label{eq:notable:rot_gamma_combo}
\Gamma^r{}_{t\varphi}+\omega\,\Gamma^r{}_{\varphi\varphi}
=
\frac{r^2\sin^2\theta}{2B}\,\omega'(r),
\end{equation}
so that, using $a_r=N'/N$ and $u^t=1/N$, $u^\varphi=\omega/N$,
\begin{equation}
\label{eq:notable:rot_dota_phi}
\dot a_\varphi
=
-\frac{r^2\sin^2\theta}{2B}\,\frac{N'(r)\,\omega'(r)}{N(r)^2}\,.
\end{equation}

Since $\dot a_r=\dot a_\theta=0$, while $\dot a_\varphi\neq 0$ the spatial projection $\dot a_{\langle a\rangle}\equiv h_a{}^{b}\dot a_b$ has only an azimuthal component,
$\dot a_{\langle\varphi\rangle}=\dot a_\varphi$, and is orthogonal (in the $h_{ab}$ sense) to $a^a$:
$a^a\dot a_{\langle a\rangle}=0$. Therefore the projector
$P_a{}^{b}$ acts trivially on $\dot a_{\langle a\rangle}$, so
$j_{\perp a}\equiv P_a{}^{b}\dot a_{\langle b\rangle}=\dot a_{\langle a\rangle}$
implying
\begin{equation}
j_{\perp\varphi}=\dot a_\varphi,\quad j_{\perp r}=j_{\perp\theta}=0
\end{equation}
that allows to rewrite the $r$ component of Eq.~\eqref{eq:curl V perp} as
\begin{align}
    \varepsilon_{rbc} D^b V_{\perp}^c &= -2X\varepsilon_{r\theta\varphi}\left(  D^{\theta} j_{\perp}^\varphi-D^\varphi j^\theta_\perp\right) \nonumber \\
    &= -4X\cos(\theta)\sqrt{B(r)}\ j^{\varphi}_{\perp}
\end{align}
where $j^{\varphi}_{\perp} = h^{\varphi\varphi} \dot{a}_\varphi=\left(\frac{1}{2B}\,\frac{N'(r)\,\omega'(r)}{N(r)^2}\right)$ and where it was used the $h_{ij}$ built connections$
{}^3\Gamma^{\varphi}{}_{\theta\varphi}=\cot\theta,
{}^3\Gamma^{\theta}{}_{\varphi\varphi}=-\sin\theta\cos\theta
$. This expression shows that $\varepsilon_{abc}D^bV_{\perp}^{\,c}\neq 0$ on any case where
$N'(r)\omega'(r)\neq 0$ and $\cos\theta\neq 0$ (i.e.\ away from the equatorial plane).
This allows to conclude that $V_{\perp a}$ is not generically irrotational in the scalar-comoving three-space,
and hence it cannot, in general, be represented as $(D_a\Psi)_\perp$ for a scalar $\Psi$.

\subsection{Spherically symmetric example}
Consider a spherically symmetric spacetime with line element
\begin{equation}
  ds^2 = -f(t,r)\,dt^2 + g(t,r)\,dr^2 + r^2 d\Omega^2,
\end{equation}
with $f(t,r)>0$, $g(t,r)>0$.
Spherical symmetry requires the scalar field to depend only on $t$ and $r$,
\begin{equation}
  \Phi = \Phi(t,r),
\end{equation}
with timelike gradient so that $\nabla_a\Phi$ cannot be purely radial. Spherical symmetry implies that any spatial vector field orthogonal to $u^a$ and built covariantly from $(g_{ab},\Phi,u^a)$ must be tangent to the unique radial direction in the comoving 3--space. Denoting by $\hat r_a$ the unit radial covector in the spatial
metric $h_{ab}$, we therefore have $a_a = a_r(t,r)\,\hat r_a, D_a\dot X = (D_r\dot X)(t,r)\,\hat r_a, \sigma_a{}^b a_b = s_r(t,r)\,\hat r_a$, for suitable scalar functions $a_r$, $D_r\dot X$, and $s_r$. Using the general expression~\eqref{eq:Va_final} it follows that $V_a$ is also purely radial
\begin{equation}
  V_a = V_r(t,r)\,\hat r_a,
\end{equation}
with $V_r$ a scalar. Whenever $a_a\neq 0$, $V_a$ is then parallel to $a_a$, since both are proportional to the same unit vector $\hat r_a$.

The component of $V_a$ orthogonal to $a_a$ is defined by $V_{\perp a} \equiv P_a{}^b V_b$, and because $V_a$ is parallel to $a_a$, the projector $P_a{}^b$ annihilates it
\begin{equation}
  V_{\perp a} = P_a{}^b V_b = 0.
\end{equation}
 
Thus, for spherically symmetric configurations with timelike scalar gradient and
$\Phi=\Phi(t,r)$, the transverse piece of the scalar heat flux vanishes identically,
\begin{equation}
  q^{(\Phi)}_a
  = -\alpha(\Phi,X,\dot X,\ldots)\,a_a,
\end{equation}
even when $F_X\neq 0$.
In this highly symmetric sector the scalar heat flux is therefore aligned with the acceleration and admits a single-direction Eckart heat-flux interpretation.

\subsection{Homogeneous FLRW example}
For a spatially flat FLRW metric
$\mathrm{d}s^2=-\mathrm{d}t^2+\mathrm{a}^2(t)\delta_{ij}dx^idx^j$
and a homogeneous scalar $\Phi(t)$, one has
\begin{equation}
  u^a=(1,0,0,0),\quad a_a=0,\quad \sigma_{ab}=0,
\end{equation}
and from $X=-\tfrac12\nabla_a\Phi\nabla^a\Phi$,
\begin{equation}
  X = \frac12\dot\Phi^2(t),
\end{equation}
so $D_i X=D_i\dot X=0$ and hence $V_i=0$.
Thus $V_{\perp i}=0$ and $q^{(\Phi)}_i=0$ even for $F_X\neq0$.
Highly symmetric backgrounds admit therefore the Eckart description trivially.

\section{Discussion and outlook}\label{sec:VI}

We have shown that, within the broad class of metric single--scalar theories with action~\eqref{eq:action}, the requirement that the scalar sector admit a Eckart heat flux~\eqref{eq:Eckart Decomposed} in the scalar–comoving frame, \emph{for all} timelike scalar configurations, is equivalent to the condition
$F_X(\Phi,X)\equiv 0$, i.e.\ $F(\Phi,X)=F(\Phi)$. In this sense, the familiar Jordan–like $F(\Phi)R+G(\Phi,X)$ models are precisely those in which the extra scalar degree of freedom behaves as an Eckart–type heat–flux carrier aligned with its own four–acceleration in the comoving frame.

Our use of Eckart theory should be understood as a phenomenological \emph{selection rule}, not as a fundamental assumption about the microphysics of the scalar. Starting from the general family $F(\Phi,X)R+G(\Phi,X)$, we impose the following purely kinematical requirement: in the scalar–comoving frame defined by $u_a\propto\nabla_a\Phi$ (with timelike gradient $X>0$), the effective scalar stress–energy $T^{(\Phi)}_{ab}$ can be interpreted as that of a single–component relativistic fluid whose heat flux satisfies the standard Eckart relation $q^{(\Phi)}_a=-K(D_aT_g+T_g a_a)$ for some temperature $T_g$ and conductivity $K>0$, for all timelike scalar configurations. Comparing this structure with the general decomposition~\eqref{eq:q_decomp}, we find that the transverse contribution $(F_X/8\pi F)V_{\perp a}$ cannot, in general, be written as a spatial temperature gradient, even allowing for a component $(D_aT_g)_\perp$ perpendicular to $a_a$. Thus a global Eckart description is possible if and only if $F_X\equiv 0$, which removes the transverse term and yields a purely accelerative heat flux of the form $q^{(\Phi)}_a=-f(\Phi,X,\dot X)\,a_a$.

The corresponding theories belong to the Horndeski subclass with $G_3=0$ and $G_{4X}=G_5=0$, with the latter conditions being imposed by gravitational–wave constraints~\cite{Ezquiaga:2017ekz,Baker:2017hug}. In this way, our result complements existing theoretical and observational arguments by adding a simple thermodynamic–consistency criterion: within the $F(\Phi,X)R+G(\Phi,X)$ family, only Jordan–like nonminimal couplings allow a single–temperature, Eckart–type fluid description of the scalar sector for arbitrary timelike configurations. Moreover, from a phenomenological standpoint, it is also timely to note that recent DESI-informed cosmological analyses have begun to highlight non-minimally coupled quintessence as a particularly competitive description of late-time acceleration. In particular, in~\cite{Wolf:2025jed} it was reported strong Bayesian evidence for a non-derivative non-minimal coupling to curvature in quintessence-like models, using combinations of distance data (including DESI BAO and supernova samples), and emphasized the implications for the effective dark-energy equation of state. This is precisely the Jordan-like sector in which our kinematical criterion for a global Eckart interpretation is satisfied: requiring a standard single-temperature Eckart heat flux for arbitrary timelike scalar configurations selects $F(\Phi,X)=F(\Phi)$, excluding kinetic non-minimal couplings with $F_X\neq 0$.

Our analysis is kinematical and first order: we do not construct an entropy current, impose a second law, or address the known causality issues of Eckart
theory. Nevertheless, the Eckart requirement already acts as a strong filter on theory space. It would be natural to extend this work by (i) embedding the transverse
vector $V_{\perp a}$ that arises for $F_X\neq 0$ in causal second–order transport frameworks, (ii) exploring more general scalar–tensor and scalar–torsion theories
where additional transport channels may appear, and (iii) studying configurations where $V_{\perp a}=D_a\Psi$ for some scalar $\Psi$, which could provide controlled examples of non–Jordan–like models with a richer, yet still tractable, effective
gravitational thermodynamics.

\begin{acknowledgments}
We thank the anonymous referee for the valuable suggestions provided, which have greatly contributed to enhancing the quality of this manuscript. The authors acknowledge funding from the Fundação para a Ciência e a Tecnologia (FCT) through the research grants UIDB/04434/2020, UIDP/04434/2020 and PTDC/FIS-AST/0054/2021. D.S.P also thanks Miguel A.S Pinto for insightful discussions that improved the quality of this work. 
\end{acknowledgments}

\bibliographystyle{apsrev4-2} 
\bibliography{biblio}

\end{document}